\documentclass{article}
\usepackage{spconf,amsmath,graphicx,booktabs,algorithm,algpseudocode,booktabs,multirow}
\usepackage{amssymb,enumitem,color}
\usepackage{marvosym}

\usepackage{float}

\newcommand{\red}[1]{\textcolor{red}{#1}}

\title{Enhancing Adversarial Training with Prior Knowledge Distillation for Robust Image Compression}
\name{Zhi Cao$^{1}$\sthanks{These authors contributed equally and are regarded as co-first authors.  \ \  \ \textsuperscript{\Letter} Corresponding\ author,\ email: liangys@hit.edu.cn}, Youneng Bao$^{1,2\, *}$, Fanyang Meng$^{2}$, Chao Li$^{1}$, Wen Tan$^{1}$, Genhong Wang$^{1}$, Yongsheng Liang$^{1,}$\textsuperscript{\Letter} }
\address{$^{1}$Harbin Institute of Technology, Shenzhen, China\ \ \ \ \ $^{2}$Peng Cheng Laboratory, China}
\begin{document}
%
\maketitle
\begin{abstract}
Deep neural network-based image compression (NIC) has achieved excellent performance, but NIC method models have been shown to be susceptible to backdoor attacks.
Adversarial training has been validated in image compression models as a common method to enhance model robustness. However, the improvement effect of adversarial training on model robustness is limited. In this paper, we propose a prior knowledge-guided adversarial training framework for image compression models.
Specifically, first, we propose a gradient regularization constraint for training robust teacher models. Subsequently, we design a knowledge distillation-based strategy to generate a priori knowledge from the teacher model to the student model for guiding adversarial training.
Experimental results show that our method improves the reconstruction quality by about 9dB when the Kodak dataset is elected as the backdoor attack object for psnr attack. Compared with Ma2023\cite{C14chen2023towards}, our method has a 5dB higher PSNR output at high bitrate points.

\end{abstract}

\section{Introduction}
\label{sec:intro}

Deep learning has undergone extensive research in the field of image compression\cite{C1balle2016end,C2balle2018variational,C3minnen2018joint,C4minnen2020channel,C5cheng2020learned,Bao}. In terms of rate-distortion performance, the performance of deep learning models has surpassed that of traditional methods such as JPEG\cite{C6wallace1991jpeg}, JPEG2000\cite{C7rabbani2002overview}, BPG\cite{C8yee2017medical}, and even the recent Versatile Video Coding (VVC)\cite{C9bross2021developments}. However, as deep learning-based image compression models are applied in practice, considering the sensitivity of deep neural networks to minor input perturbations, the robustness of these models against adversarial perturbations has become an essential and unavoidable concern.

\begin{figure}[t]
    \centering
    \includegraphics[width=8.5cm]{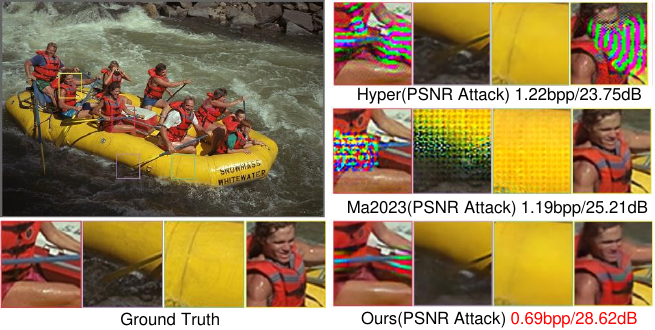}
    \caption{Performance of the original model, Ma2023\cite{C14chen2023towards}, and our model under attack. Our model outperforms both Hyper model and Ma2023\cite{C14chen2023towards} model in terms of both bpp and psnr performance.}
    \vspace{-1em}
\end{figure}

Since the image compression models used in practice are mostly publicly accessible, in this paper, we assume that attackers can access all parameters of the neural network compression model. Therefore, the attacks mentioned in this paper are all considered white-box attacks.

The concept of adversarial samples was first introduced by Szegedy\cite{C10szegedy2013intriguing}, which unveiled the sensitivity of deep neural networks to input perturbations. Following this, Goodfellow introduced an efficient attack method called the Fast Gradient Sign Method (FGSM) by utilizing gradient signs\cite{C11goodfellow2014explaining}. Kurakin enhances the attack capability by breaking down the larger step in the FGSM iteration into many smaller steps\cite{C12kurakin2018adversarial}. Madry proposed the Projected Gradient Descent (PGD) method, which increases attack success rates through multiple iterations\cite{C13madry2017towards}. In the context of image compression tasks, Ma introduced a trainable noise as an input perturbation\cite{C14chen2023towards}, using this noise during training to update the model's input and generate highly potent adversarial samples. Liu leveraged the PGD method to efficiently attack the model's bpp (bits per pixel) parameter\cite{C15liu2022denial}. We have enhanced the techniques proposed in \cite{C14chen2023towards}, introducing two attack methods tailored for image compression:  \textit{psnr attack} and \textit{bpp attack}.
\begin{figure}[h]
    \centering
    \includegraphics[width=8cm]{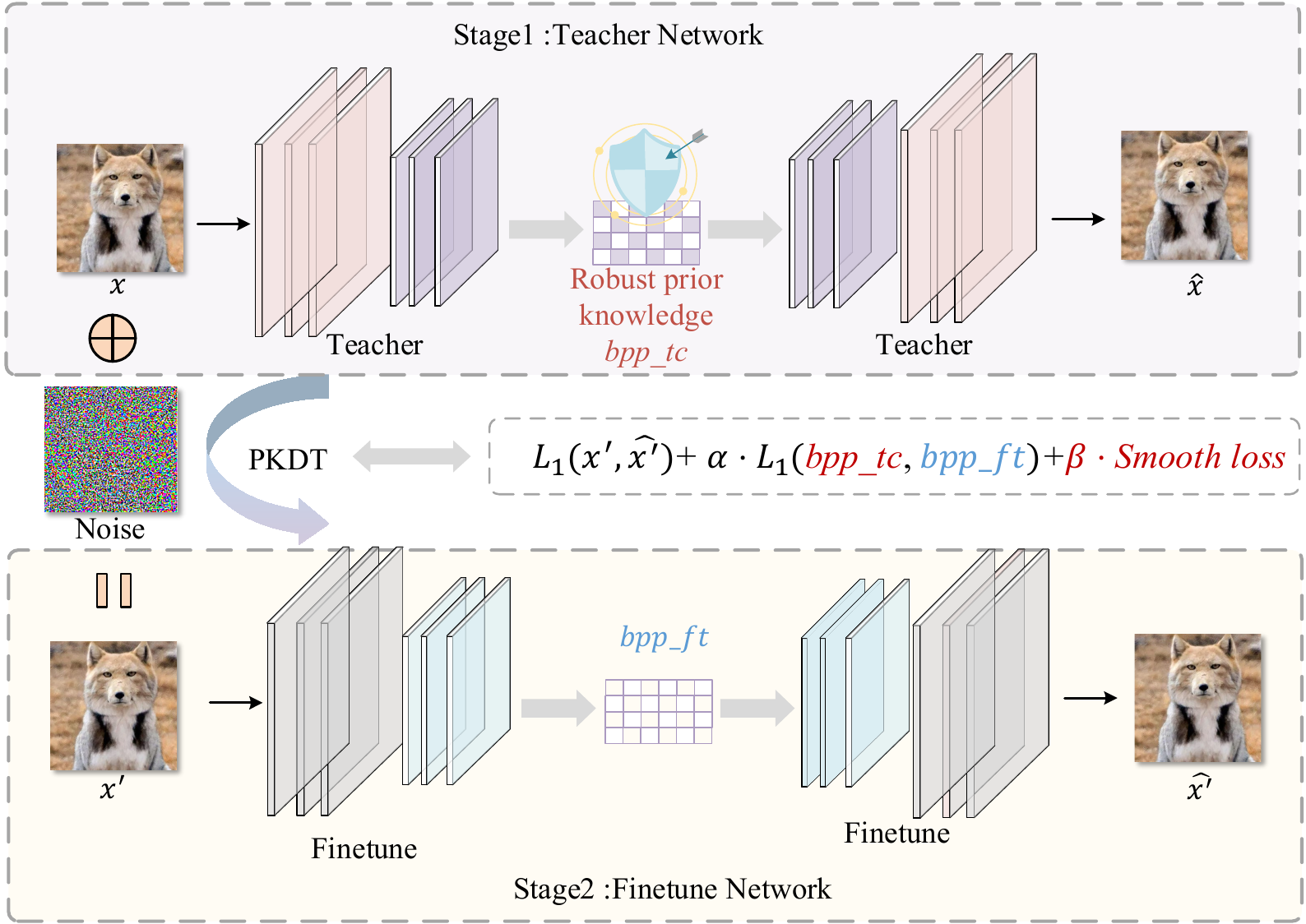}
    \caption{ Prior knowledge-guided adversarial training framework for image compression models. In the first stage, gradient regularization constraints are used to train robust teacher models. Subsequently, in the second stage, a prior knowledge distillation transfer strategy (PKDT) is used to transfer prior knowledge generated by the teacher model to the finetune model to guide adversarial training. }
    \vspace{-1em}
    \label{fig:model}
\end{figure}

Adversarial Training\cite{C16goodfellow2014explaining,C17madry2017towards}, serves as the most effective method to enhance a model's resilience against adversarial attacks. The core idea of this technique is to introduce adversarial samples and use them as supplementary data alongside the original dataset for model training. The training process can be framed as a min-max problem, where the maximization problem involves generating adversarial samples with strong attack capabilities and the minimization problem focuses on training the model to minimize the output loss for these adversarial samples. However, applying adversarial training directly to image compression tasks resulted in relatively minor improvements in model robustness. To further enhance the model's robustness, we introduce prior knowledge into the adversarial training process of the model through a distillation framework\cite{C20papernot2016distillation}. Contributions of this paper are summarized as follows:

$\bullet$ We propose a $smooth\_loss$ to train the teacher model and utilize this teacher model to generate strong robustness-oriented prior knowledge.

$\bullet$ We propose a distillation framework to transfer prior knowledge and use this knowledge to guide the model's adversarial training.

$\bullet$ Extensive experiments demonstrate that our proposed method significantly enhances the model's robustness. In the Kodak dataset, our method can achieve a maximum robustness improvement of up to 23\% against \textit{psnr attack}.

\begin{algorithm}[ht]
	\caption{Adversarial Training with Prior Knowledge}   
	\begin{algorithmic}[1] 
	  \Require Pre-trained model $M\_pre$, Dataset 
         $D$, Training epoch $N$, \textit{psnr attack} iterations $N_{p}$, \textit{bpp attack } iterations $N_{b}$
        \Ensure  Finetune model $M\_ft$   
        \State \textbf{Stage} \uppercase\expandafter{\romannumeral1}: Training a robustness teacher model $M\_tc$
        \State Let $M\_tc=M\_pre$
        \For {$i<N$}
            \State $\vec{x}=Random\_Sample(D,batchsize=16)$
            \State $smooth\_loss = \frac{{\partial {\rm{bpp}}\_main}}{{\partial x}} + \frac{{\partial y}}{{\partial x}}$
            \State $loss=RD\_loss+ \alpha \cdot smooth\_loss$
            \State updating the parameters of model $M\_tc$ by $loss$
        \EndFor
        \State return teacher model $M\_tc$
        \State \textbf{Stage} \uppercase\expandafter{\romannumeral2}: Finetune network, adversarial training with prior knowledge
        \State Let $M\_ft=M\_pre$
        \For {$i< N$}
            \State $\vec{x}_{ori}=[x_{1},x_{2},x_{3}]$
            \Statex$\quad\quad\quad=Random\_Sample(D,batchsize=3)$
            \State Generate adversarial sample $\acute{x_{1}}$ through \textit{psnr attack} 
            \Statex \quad \ \ on $x_{1}$ with $N_{p}=200$, generate adversarial sample
            \Statex \quad \ \ $\acute{x_{2}}$ through \textit{bpp attack} on $x_{2}$ with $N_{b}=100$
            \State $\vec{x}_{new}= [{x}'_{1},{x}'_{2},x_{3}]$
            \State Input $\vec{x}_{ori}$ into model $M\_tc$ to obtain $bpp\_tc$
            \Statex \quad \ \ Input $\vec{x}_{new}$ into model $M\_ft$ to obtain $bpp\_ft$
            \State $D\_loss = {\left\| {{{\vec x}_{new}} - M\_ft({{\vec x}_{new}})} \right\|_1}$
            \State $R\_loss = {\left\| {bpp\_tc - bpp\_ft} \right\|_1}$
            \State $smooth\_loss = \frac{{\partial {\rm{bpp}}\_main}}{{\partial x}} + \frac{{\partial y}}{{\partial x}}$
            \State $loss = D\_loss + \alpha  \cdot R\_loss +  \beta  \cdot smooth\_loss$
            \State updating the parameters of model $M\_ft$ by $loss$
        \EndFor
   \end{algorithmic} 
\end{algorithm} 

\section{Method}
\label{sec:format}
Our method effectively improves the model's robustness by introducing a distillation framework for transferring prior knowledge. The framework of our model is illustrated in Fig.\red{\ref{fig:model}} and consists of two components: a teacher network and a finetune network. The teacher network provides the prior knowledge necessary for training the finetune network. A gradient regularization term be used to train the teacher network, enabling it to generate prior knowledge with robustness-related information. The finetune network involves using prior knowledge generated by the teacher model to guide adversarial training. 
Since the $psnr$ error caused by attacks is relatively large, using $psnr$ as prior knowledge would make the adversarial training process difficult to converge. At the same time, the error of $bpp$ is very small, so we choose $bpp$ as the prior knowledge.
\subsection{Stage \uppercase\expandafter{\romannumeral1}: Training a Robustness Teacher}
Gradient regularization has been demonstrated as an effective method for improving model robustness\cite{C18drucker1992improving,C19finlay2021scaleable}. However, due to the presence of two performance metrics for the image compression task, we have designed two gradient constraint terms. Regarding the $bpp$ term, we observed that the changes in $bpp$ after the sample is attacked are mainly manifested in the $bpp$ during the analysis transformation process, which we refer to as $bpp\_main$.  Therefore, the constraint term is formulated as $\partial {\rm{bpp}}\_main / \partial x$. For the $psnr$ term, we also utilize the gradient of $y$ with respect to $x$ after the analysis transformation as another constraint. Therefore, the constraint term for gradient smoothness is formulated as:
\begin{equation}
    smooth\_loss = \underbrace{\frac{{\partial {\rm{bpp}}\_main}}{{\partial x}}}_{bpp \ constrain} + \underbrace{\frac{{\partial y}}{{\partial x}}}_{psnr \ constrain}
\end{equation}
Then the overall training loss of the model is:
\begin{equation}
    loss = RD\_loss + smooth\_loss
\end{equation}
where $RD\_loss$ is the rate-distortion loss function of the \textit{pre-trained} model, and $RD\_loss=R+\lambda D$, Distortion term $D$ is measured by peak signal-to-noise ratio (PSNR) between original images and reconstructed images, $R$ is the bit rate, $\lambda$ is hyper parameter to control the trade-off between rate and distortion. Through training, we obtain the teacher model, denoted as \textit{gradient} model.

\subsection{Stage \uppercase\expandafter{\romannumeral2}: Transferring Prior Knowledge}
We transfer prior knowledge from the teacher network to the finetune network through a distillation framework. The prior knowledge specifically refers to the $bpp$ generated by input samples into the teacher model, denoted as $bpp\_tc$, We assume the input sample as $x$, and the reconstructed output from the sample input into the finetuned model as $\hat{x}$, with $bpp$ as $bpp\_ft$. The loss for training the finetune network consists of three components. First is the reconstruction error part:
\begin{equation}
    D\_loss = \left \| x-\hat{x}  \right \| _{1}
\end{equation}
Secondly, there is the introduction of prior knowledge:
\begin{equation}
    R\_loss=\left \| bpp\_tc-bpp\_ft  \right \| _{1}
\end{equation}
Lastly, the $smooth\_loss$ in Equation 2.
Then the overall training loss of the model is:
\begin{equation}
    loss=D\_loss+\alpha  \cdot R\_loss + \beta  \cdot smooth\_loss
\end{equation}
The entire training process is depicted as Algorithm 1.

\subsection{Attack Method}
The image compression task involves two evaluation metrics, $psnr$ (Peak Signal-to-Noise Ratio) and $bpp$ (Bits Per Pixel), which directly impact the model's compression performance. Therefore, we conduct attacks on these two metrics separately. When attacking $psnr$, we ensure that $bpp$ remains almost unchanged to simulate a real-world scenario where image transmission efficiency remains constant, but image decompression quality significantly deteriorates. Conversely, when attacking $bpp$, we ensure that $psnr$ remains almost unchanged to simulate a situation in the real world where image decompression quality remains constant, but transmission efficiency experiences a significant drop.

Assuming the input image is $x$, noise is $n$, adversarial samples $x^{\ast } =x+n$. The input image and the adversarial sample are both reconstructed to $\hat{x}$ and $\hat{x}^{\ast }$, simultaneously, the bit rates are $bpp\_ori$ and $bpp$, respectively. We improved the approach outlined in \cite{C14chen2023towards} by modifying the loss function to:
\begin{equation}
arg\min_{n} L_{d}=\left\{\begin{matrix}\left \|  n\right \| _{2} +bpp\_error \quad \quad \left \| n \right \|_{2}^{2}\ge \epsilon     
 \\
1-\left \| x-\hat{x}^{\ast }   \right \| _{2}^{2} \quad \quad \ \ \ \ \ \    \left \| n \right \| _{2}^{2}< \epsilon   
\end{matrix}\right. 
\end{equation}
\begin{equation}
arg\min_{n} L_{d}=\left\{\begin{matrix}\left \|  n\right \| _{2} +psnr\_error \quad \left \| n \right \|_{2}^{2}\ge \epsilon     
 \\
-bpp \quad \quad \quad \quad \quad \quad \quad \left \| n \right \| _{2}^{2}< \epsilon   
\end{matrix}\right. 
\end{equation}
where $\epsilon$ is the $l_{2} $ threshold of the input noise, $bpp\_error=\left \| bpp-bpp\_ori \right \| _{1}$, $psnr\_error=\left \| \left \| x-\hat{x} ^{\ast }  \right \|_{2}^{2}-\left \| x-\hat{x}  \right \| _{2}^{2}    \right \| _{1}$. Then implementing \textit{psnr attack} through training Equation 6, and implementing \textit{bpp attack} through training Equation 7.

\section{Experiment}
\label{sec:pagestyle}

\begin{figure*}[t]
    \centering
    \includegraphics[width=17.5cm]{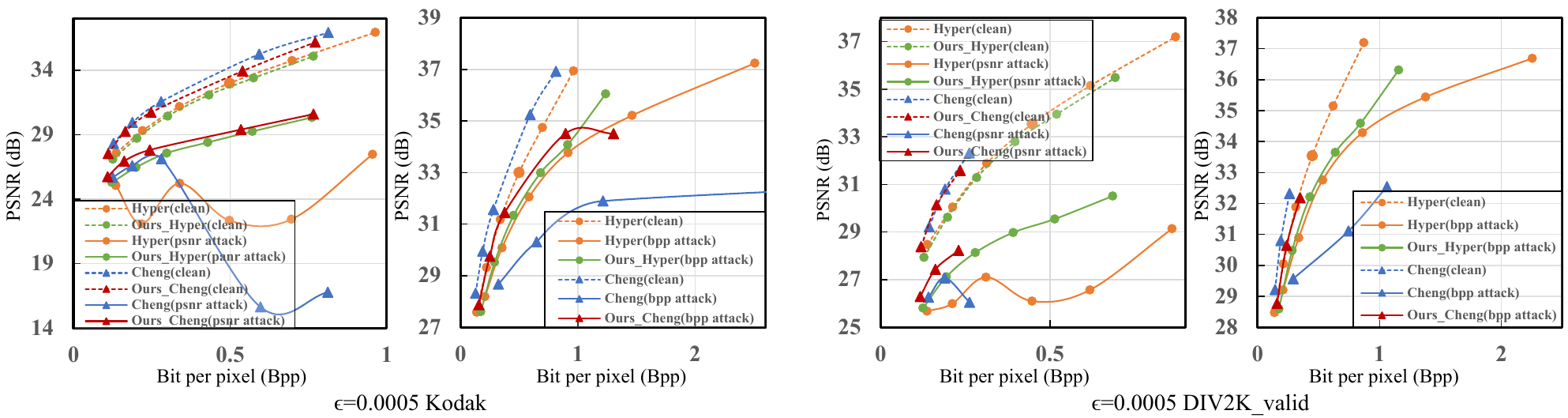}
     \vspace{-1em}
    \caption{R-D performance of the model under different attack scenarios. The two graphs on the left illustrate the robustness performance of our method applied to the \textit{Hyper} and \textit{Cheng} models using the Kodak test dataset. The first graph on the left pertains to \textit{psnr attack}, while the second one is related to \textit{bpp attack}, with the attack iterations $N_{p}=N_{b}=1000$. The two graphs on the right are based on the DIV2K valid dataset, with the attack iterations $N_{p}=N_{b}=500$. It's worth noting that in the second graph on the left, the last two data points on curve \textit{Ours\_Cheng(bpp attack)} have bpp exceeding 9.}
    \label{fig:comprison1}
    \vspace{-1em}
\end{figure*}

\begin{figure}[htbp]
    \centering
    \includegraphics[width=8.5cm]{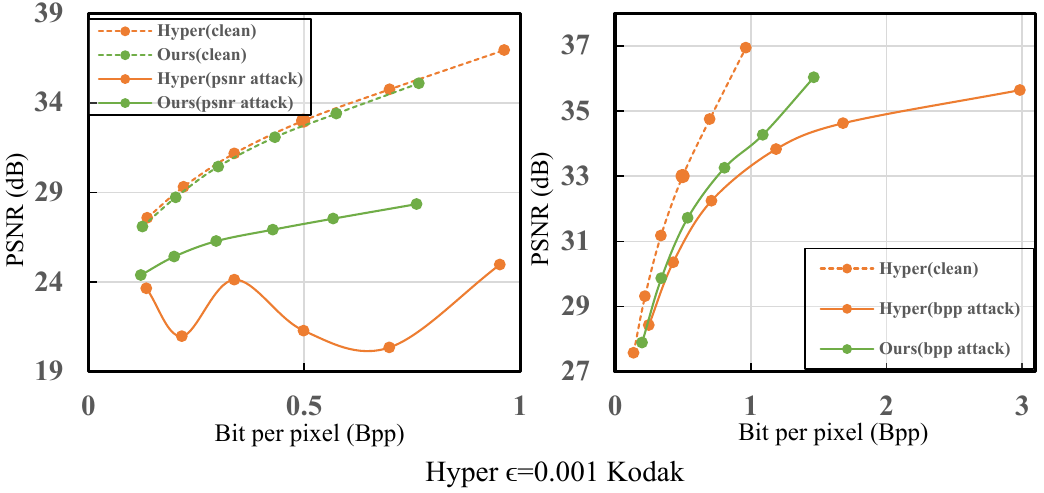}
    \caption{R-D performance of the model under other $\epsilon$}
    \label{fig:epsilon}
    \vspace{-1em}
\end{figure}

$A.$ \textit{Experiment Setup}

\textbf{Datasets.} The training dataset includes CLIC and DIV2K. Two datasets are used for testing, including Kodak dataset with 24 images at a size of 768×512, DIV2K valid dataset with 100 high-quality images.

\textbf{Training details.} We train each model for 100 epochs with Adam optimizer. For the \textit{pre-train model}, the learning rate is initially set to $10^{-4}$ and then halved every 25 epochs after the initial 50 epochs. For the \textit{gradient model} and ours finetune model the learning rate is initially set to $5\times 10^{-5}$, and then halved every 25 epochs after the initial 25 epochs. The \textit{pre-trained model} is trained using the CLIC dataset. The \textit{gradient model} and our fine-tuned model are trained using the DIVK dataset, more importantly, both of them are fine-tuned based on the \textit{pre-trained model}. In the adversarial training process we proposed, the number of iterations for \textit{psnr attack} $N_{p} $ is set to 200, while for \textit{bpp attack}, iteration counts $N_{b}$ is set to 100, and $\epsilon=0.0005$.\\
$B.$ \textit{Results and Comparison}

We tested two original models, \textit{Hyper}\cite{C2balle2018variational} and \textit{Cheng}\cite{C5cheng2020learned}, as well as the finetuned models obtained using our method. We evaluated these models on various test datasets and under different attack methods to obtain results. Fig.\red{\ref{fig:comprison1}} shows the performance comparison of our model in different scenarios. It can be observed that when our method is applied to different models, there is a noticeable enhancement in robustness. Especially for models with large parameter sizes like \textit{Cheng} the improvement in robustness is more pronounced compared to the \textit{Hyper} model. Additionally, we can see that our method effectively enhances model robustness in both low-resolution and high-resolution test datasets. Fig.\red{\ref{fig:epsilon}} illustrates the robustness performance of our method concerning different noise thresholds for $\epsilon=0.001$. It can be observed that the model's robustness is significantly enhanced when compared to smaller noise thresholds $\epsilon=0.0005$. Particularly, in the case of \textit{bpp attack}, the robustness improvement can reach up to $23\%$. Table.\red{\ref{table:1}} contains the data points for the low bitrate scenario shown in Fig.\red{\ref{fig:comprison1}}, leftmost graph. It can be observed that the robustness improvement brought by our method is similarly higher than that of the \textit{gradient} model, which is training with stage \uppercase\expandafter{\romannumeral1}.

\begin{table}[h]
\centering
\caption{R-D performance of model}
\resizebox{\linewidth}{!}{
\begin{tabular}{c c c c c c c}
\toprule
\multirow{2}*{model}&\multicolumn{2}{c}{clean}&\multicolumn{2}{c}{psnr attack}&\multicolumn{2}{c}{bpp attack}\\
\cmidrule(r){2-3}  \cmidrule(r){4-5} \cmidrule(r){6-7} 
~&bpp&psnr&bpp&psnr&bpp&psnr \\
\midrule
Hyper&0.4978 &33 &0.498 &22.3502 &0.9161 &33.7702\\
Ours w/o Stage \uppercase\expandafter{\romannumeral2}&0.4322 &32.2592 &0.4309 &25.9039 &0.7225 &33.1432\\
Ours&0.4323 &32.0802 &\red{0.4286} &\red{28.4039} &\red{0.6822} &32.9933\\
\bottomrule
\end{tabular}}
\label{table:1}
\vspace{-1em}
\end{table}

\begin{figure}[H] 
    \centering
    \includegraphics[width=7cm]{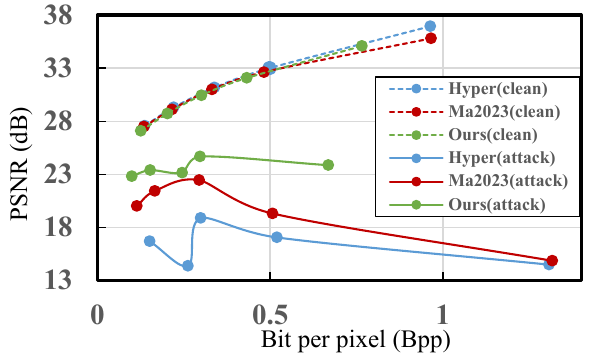}
    \caption{R-D performance compared with Ma2023}
    \label{fig:Ma2023}
    \vspace{-1em}
\end{figure}

Adversarial training can potentially lead to a decrease in the model's generalization ability, resulting in less improvement in robustness against other attack methods. However, we compared our approach with the attack method mentioned in Ma2023. The results of the comparison can be seen in Fig.\red{\ref{fig:Ma2023}}. It is evident that for adversarial samples where we did not undergo adversarial training, our method still exhibits better robustness.
\noindent $C.$ \textit{ Ablation Experiments}

We demonstrate the optimality of our method under the same training parameter conditions in Table.\red{\ref{table:ablation}}. The second row represents the most effective method used in Algorithm 1. The third row represents the results obtained by training without the $smooth\_loss$ term in the stage \uppercase\expandafter{\romannumeral2}. Analogously, the results in the last two rows correspond to the removal of $\partial y/\partial x$ and $\partial {\rm{bpp}}\_main / \partial x$ in stage \uppercase\expandafter{\romannumeral2}, respectively. Test dataset is Kodak with iteration counts $N_{p}=400$, $N_{b}=200$, and noise threshold $\epsilon=0.0005$. From the results, we can observe that removing any of the constraint terms deteriorates the model's robustness. Additionally, we have noticed that the impact on robustness is relatively small. This further validates that in 
our proposed method, the most significant improvement in model robustness is achieved through the introduction of prior knowledge.
 

\begin{table}[h]
\centering
\caption{Results of the ablation experiments}
\resizebox{\linewidth}{!}{
\begin{tabular}{c c c c c c c}
\toprule
\multirow{2}*{model}&\multicolumn{2}{c}{clean}&\multicolumn{2}{c}{psnr attack}&\multicolumn{2}{c}{bpp attack}\\
\cmidrule(r){2-3}  \cmidrule(r){4-5} \cmidrule(r){6-7} 
~&bpp&psnr&bpp&psnr&bpp&psnr \\
\midrule
best&0.4323 &32.0802 &\red{0.4273} &\red{28.6556} &\red{0.6953} &33.0194\\
\cmidrule(r){1-1}
best w/o \\$smooth\_loss$&0.4358 &32.1936 &0.4328 &28.4497 &0.7214 &33.0731\\
\cmidrule(r){1-1}
best w/o \\$\partial y/\partial x$&0.4355 &32.1549 &0.4301 &28.5293 &0.7075 &33.031\\
\cmidrule(r){1-1}
best w/o \\$\partial {\rm{bpp}}\_main/\partial x$&0.4332 &32.1463 &0.4279 &28.6224 &0.7083 &33.093\\
\bottomrule
\end{tabular}}
\label{table:ablation}
\vspace{-1em}
\end{table}


\section{conclusion}
\label{sec:page}
In this paper, we propose a method of training the teacher model using gradient regularization and use the prior knowledge generated by this teacher model to guide the entire process of adversarial training. Experimental results demonstrate that our approach can effectively improve the robustness of the model. This is especially effective when dealing with large-parameter models and large noise thresholds.


\bibliographystyle{ieeetr}
\bibliography{refs} 

\end{document}